\documentclass{article}

\usepackage{arxiv}
\usepackage[utf8]{inputenc} 
\usepackage[T1]{fontenc}    
\usepackage{url}            
\usepackage{booktabs}       
\usepackage{amsfonts}       
\usepackage{nicefrac}       
\usepackage{microtype}      
\usepackage{lipsum}
\usepackage{subcaption}
\usepackage{graphicx}
\graphicspath{ {./images/} }
\usepackage{graphicx}
\usepackage{amsmath}
\usepackage{amssymb}
\usepackage{booktabs}
\usepackage{float}
\usepackage{multirow}

\usepackage[linesnumbered,ruled,vlined]{algorithm2e}
\usepackage{algpseudocode}

\usepackage[pagebackref,breaklinks,colorlinks]{hyperref}
\usepackage[capitalize]{cleveref}

\usepackage{cite}
\usepackage{textcomp}
\usepackage{xcolor}

\title{
PTQ4ADM: Post-Training Quantization for Efficient Text Conditional Audio Diffusion Models
}

\author{
Jayneel Vora$^{1}$, Aditya Krishnan$^{2}$, Nader Bouacida$^{1}$, Prabhu RV Shankar$^{3}$,  Prasant Mohapatra$^{4}$\\
$^{1}$Department of Computer Science, University of California, Davis\\
$^{2}$Department of Electrical and Computer Engineering, University of California, Davis\\
$^{3}$Department of Public Health Sciences, University of California Davis Health\\
$^{4}$Department of Computer Science, University of South Florida,Tampa\\
\texttt{$^{1}$\{jrvora,nbouacida\}@ucdavis.edu}\\
\texttt{$^{2}$adikrishnan@ucdavis.edu}\\
\texttt{$^{3}$rvpshankar@ucdavis.edu}\\
\texttt{$^{4}$pmohapatra@usf.edu}
}

\begin{document}
\maketitle

\begin{abstract} 

Denoising diffusion models have emerged as state-of-the-art in generative tasks across image, audio, and video domains, producing high-quality, diverse, and contextually relevant data. However, their broader adoption is limited by high computational costs and large memory footprints. Post-training quantization (PTQ) offers a promising approach to mitigate these challenges by reducing model complexity through low-bandwidth parameters. 
Yet, direct application of PTQ to diffusion models can degrade synthesis quality due to accumulated quantization noise across multiple denoising steps, particularly in conditional tasks like text-to-audio synthesis.
This work introduces PTQ4ADM, a novel framework for quantizing audio diffusion models(ADMs). Our key contributions include (1) a coverage-driven prompt augmentation method and (2) an activation-aware calibration set generation algorithm for text-conditional ADMs. These techniques ensure comprehensive coverage of audio aspects and modalities while preserving synthesis fidelity. We validate our approach on TANGO, Make-An-Audio, and AudioLDM models for text-conditional audio generation.
Extensive experiments demonstrate PTQ4ADM's capability to reduce the model size by up to 70\% while achieving synthesis quality metrics comparable to full-precision models($<$5\% increase in FD scores). We show that specific layers in the backbone network can be quantized to 4-bit weights and 8-bit activations without significant quality loss. This work paves the way for more efficient deployment of ADMs in resource-constrained environments.

\end{abstract}


\section{Introduction}
Audio Diffusion Models (ADMs) have shown significant success in generating high-quality audio across various modalities using iterative diffusion-denoising processes \cite{liu2023audioldm, huang2023make, majumder2024tango}. 
These models produce audio with realistic temporal and spectral characteristics, which is crucial for applications like music production \cite{mittal2021symbolic}, sound design \cite{suckrow2024diffusion}, and speech synthesis \cite{huang2022fastdiff}.

ADMs offer detailed control over both temporal structure and frequency content, enabling adoption in industry systems like Google's MusicLM \cite{agostinelli2023musiclm} and ByteDance's MeLoDy \cite{lam2024efficient}, which are based on latent diffusion models. 
Both are industry standards for music generation, audio editing, and voice cloning, showcasing the transformative potential of ADMs in audio technology.

However, due to their memory footprint and iterative sampling, ADMs are computationally expensive, posing challenges for deployment in resource-constrained environments. 
This growing demand for efficient on-device audio synthesis in mobile applications, gaming, and augmented reality calls for more computationally efficient ADMs. 
One promising approach to reducing ADM synthesis costs is post-training quantization (PTQ), which compresses models without retraining \cite{shang2023post,li2023q,he2024ptqd}. 
However, applying PTQ to ADMs introduces challenges as quantization errors accumulate during the iterative process, potentially degrading audio fidelity.
This research proposes a PTQ strategy for ADMs, optimizing computational efficiency while preserving synthesis quality.

The key contributions of this paper can be summarized as follows: (1) We introduce a GPT-based Caption Coverage Module for generating diverse and representative calibration sets. (2) We propose a novel activation-aware calibration sampling algorithm that prioritizes sampling intermediates from critical timesteps in the diffusion process and creates the calibration set. (3) We evaluate our approach across multiple state-of-the-art ADMs using standardized synthesis quality metrics.

\section{Related Work}
\textbf{Text-Guided Audio Synthesis:} Text-conditioned audio synthesis has recently progressed through various methods. 
DiffSound \cite{yang2023diffsound} utilizes a diffusion model to convert mel-spectrograms into discrete codes, guided by a CLIP-based text encoder \cite{hafner2021clip}. 
AudioLM \cite{borsos2023audiolm} uses hierarchical tokenization to distinguish between semantic and acoustic tokens, allowing coherent audio generation without symbolic representations. 
AudioGen \cite{kreuk2022audiogen} adopts an autoregressive model with a Transformer-based decoder for text-conditioned audio, though it risks oversimplification by discarding important text details. 
Tango \cite{ghosal2023text} bypasses the need for a text-audio encoder like CLAP, leveraging a pre-trained language model for text guidance. 
Recent works, such as AudioLDM \cite{liu2023audioldm} and Make-an-Audio \cite{huang2023make}, explore continuous latent spaces, generating mel-spectrograms from text and synthesizing audio via HiFi-GAN. 

\textbf{Diffusion Model Compression:} Due to the high computational cost of diffusion models, 
compression methods like efficient sampling \cite{watson2021learning}, structural pruning \cite{castells2024ld}, and knowledge distillation \cite{huang2024knowledge} have been explored. 
PTQ4DM \cite{shang2023post} and Q-Diffusion \cite{li2023q} introduced post-training quantization (PTQ) for diffusion models, with PTQ4DM uniformly sampling from the denoising process and Q-Diffusion segmenting it into blocks for sampling at different timesteps. 
PTQD \cite{he2024ptqd} improves performance by disentangling quantization noise. 
Temporal dynamics have also been considered, as seen in TDQ \cite{so2024temporal}, where small neural networks predict scaling parameters. 
Enhanced quantization techniques like EDA-DM \cite{liu2024enhanced} calibrate quantized networks using representative timesteps. 
QuEST \cite{wang2024quest} emphasizes temporal activation outputs, further enhancing temporal information in quantization.

\begin{figure}[H]
    \centering
    \includegraphics[width=0.8\columnwidth]{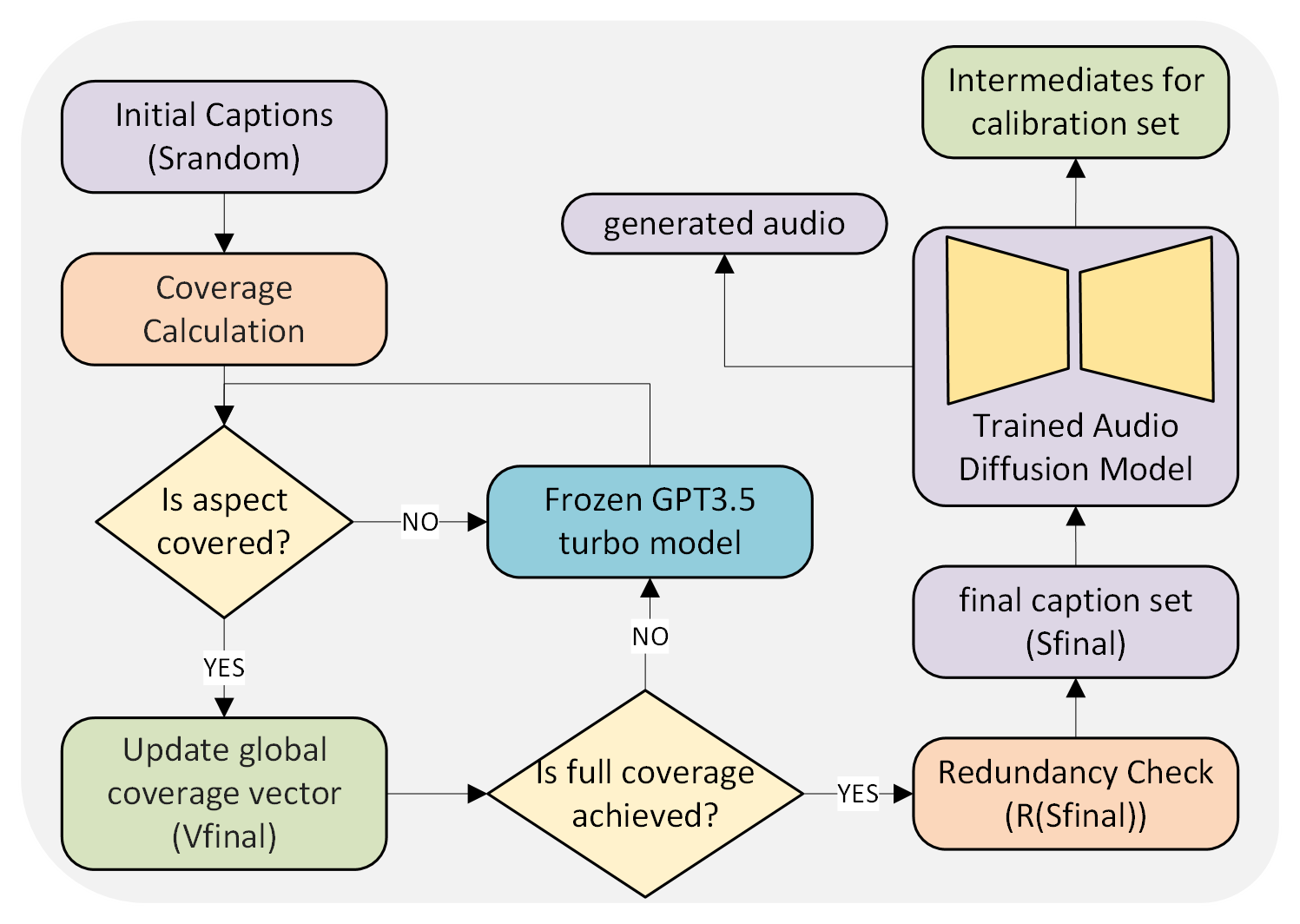}
    \caption{A diagrammatical representation of the generation of the intermediate for the calibration set using the PTQ4ADM framework.}
    \label{fig:formulation}
\end{figure}

\section{Background}
Diffusion models (DMs) \cite{sohl2015deep, ho2020denoising} consist of forward and reverse processes. In the forward process, Gaussian noise is iteratively added to data \( x_0 \) over \( T \) steps:
\begin{equation}
q(x_t | x_0) = \mathcal{N}(x_t; \sqrt{\alpha_t} x_0, (1 - \alpha_t) \mathbf{I}),
\end{equation}
where \( \alpha_t \) is the noise schedule parameter. The reverse process learns a parameterized model \( p_\theta(x_{t-1} | x_t) \) to denoise the data.

Latent diffusion models (LDMs) \cite{rombach2022high} apply this process in a latent space \( z_t \), adding noise to latent variables:
\begin{equation}
q(z_t | z_0) = \mathcal{N}(z_t; \sqrt{\alpha_t} z_0, (1 - \alpha_t) \mathbf{I}),
\end{equation}
and learning \( p_\theta(z_{t-1} | z_t) \) to reconstruct data more efficiently in compressed spaces.

Post-training quantization (PTQ) \cite{li2023q} reduces model size by quantizing weights \( w \) and activations into discrete values:
\begin{equation}
\hat{w} = s \cdot \left( \text{clip}\left( \text{round}\left(\frac{w}{s}\right) + Z, c_{\text{min}}, c_{\text{max}} \right) - Z \right),
\end{equation}
where \( s \) is the scale factor, \( Z \) is the zero-point offset, and \( c_{\text{min}}, c_{\text{max}} \) define the quantization bounds.

\section{Sensitivity Analysis}
\begin{figure}[]
    \centering
    \includegraphics[width=0.8\columnwidth]{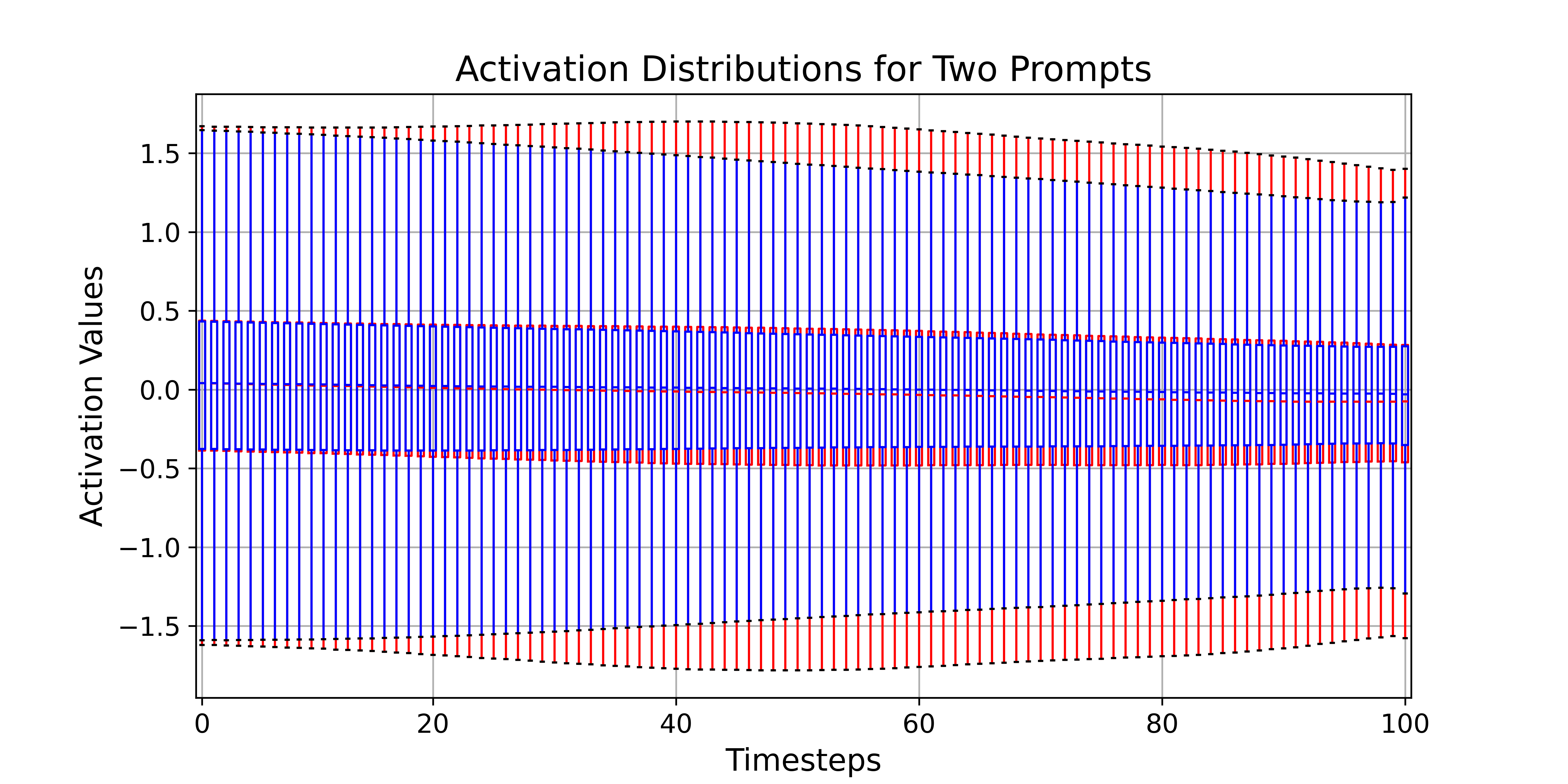}
    \caption{Activation distributions within the TimestepEmbedSequential block, located in the input layers of the Make-An-Audio diffusion model, for two distinct prompts: (1) a single, continuous alarm beep, and (2) a surreal, dissonant melody characterized by mechanical grinding, sporadic bursts of static, and alien-like vocalizations.
}
    \label{fig:activation_comparison}
\end{figure}
\begin{figure}[]
    \centering
    \includegraphics[width=0.8\columnwidth]{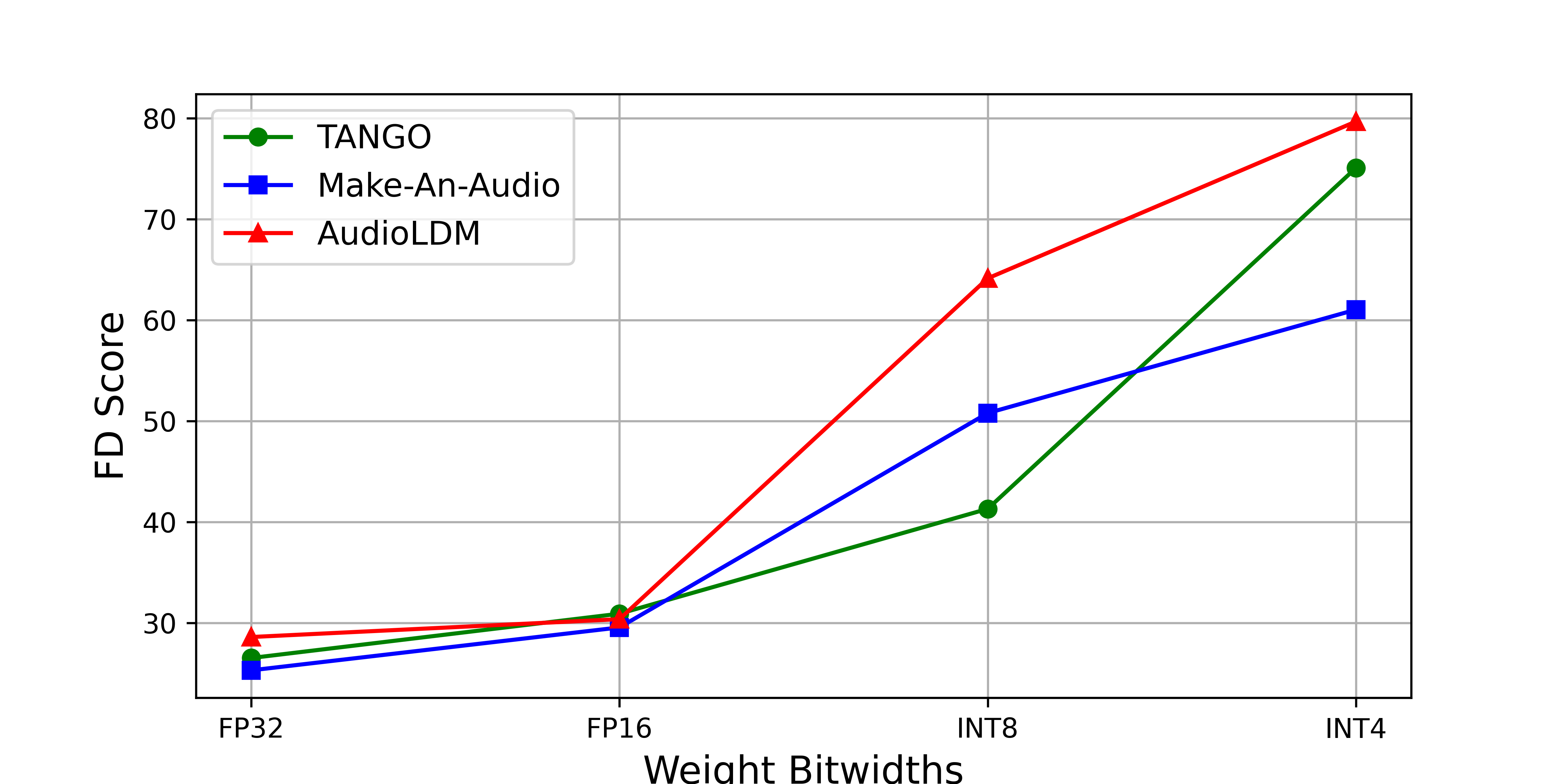}
    \caption{ FAD score for varying bitwidths for the Conv2d layer of the U-Net using uniform quantization without calibration across considered ADMs}
    \label{fig:sensitivity_weights}
\end{figure}
We conduct a sensitivity analysis to examine the activation ranges of diffusion models across various audio synthesis modalities and aspects. 
These models are expected to exhibit diverse behaviors depending on the characteristics of the input prompts. 
Figure \ref{fig:activation_comparison} compares the activation distributions for two contrasting prompts. 
The observed variation in activation ranges highlights the need for a diverse calibration set built from text captions that cover different audio characteristics and generation paths.
We identify key audio modalities such as speech, music, environmental sounds, sound effects, human vocalizations, animal sounds, and abstract sounds. 
Additionally, important aspects like temporal coherence, spatial coherence, loudness, rhythmic structure, pitch, texture, harmonicity, reverberation, and articulation are essential characteristics of the audio outputs that the diffusion models will generate and must be accounted for during calibration.
Our analysis explores the effect of uniformly quantizing the weights in convolutional and fully connected layers to different bitwidths while keeping activations in full precision. 
We measure the quality of audio (FD score) generated for the captions in the test set of AudioCaps.
As illustrated in Figure \ref{fig:sensitivity_weights}, we observe a significant drop in quality as the bitwidth decreases. 
This drop in quality emphasizes the critical importance of proper calibration in preserving model performance under low-bitwidth quantization.
Our findings on which layers to quantize align with prior research \cite{shang2023post}, reinforcing the need for careful quantization of the convolutional and fully connected layers 
while preserving the more sensitive layers in FP16 to maintain fidelity in audio synthesis.

\section{PTQ4ADM}
\label{sec:prompt_optimization}
The PTQ4ADM quantization framework comprises a coverage-driven prompt augmentation method and a novel calibration set generation method.

\subsection{Coverage-Driven Prompt Augmentation}
The coverage-driven prompt augmentation method is designed to generate a set of text captions, denoted as $S_{\text{final}}$, that fully covers a user-defined set of audio aspects, $B_{\text{custom}}$ before any audio generation or latent extraction occurs. 
This method ensures comprehensive coverage of all required aspects while maintaining diversity in the final set of captions by employing redundancy reduction techniques. We begin by initializing a set of randomly sampled captions, $S_{\text{random}}$, from the evaluation dataset or user-provided. The goal is to iteratively augment this set until it transforms into $S_{\text{final}}$, a set of captions that covers every aspect specified in $B_{\text{custom}}$.

For each caption $t_i \in S_{\text{random}}$, we assign a binary coverage vector $V_{t_i} \in \{0,1\}^{|B_{\text{custom}}|}$, where $V_{t_i}[b] = 1$ if the caption sufficiently addresses aspect $b \in B_{\text{custom}}$. 
This is determined through topic modeling. 
The global coverage vector $V_{\text{final}}$ is then defined as $V_{\text{final}}[b] = \max_{t_i \in S_{\text{final}}} V_{t_i}[b]$, ensuring that as long as at least one caption covers an aspect, it is marked as covered in the global vector. 
If any aspect $b \in B_{\text{custom}}$ remains uncovered (i.e., $V_{\text{final}}[b] = 0$), the GPT-3.5-turbo model is invoked to generate a new caption $t_j$ that targets the uncovered aspect. 
The generated caption is validated by checking its coverage vector $V_{t_j}$, and if successful, it is added to $S_{\text{final}}$ and $V_{\text{final}}$ is updated.

To ensure that the final set remains diverse and free of redundancy, a redundancy score $R(S_{\text{final}}) = \sum_{i \neq j} \mathbb{I}(V_{t_i} = V_{t_j})$ is computed, where $\mathbb{I}(\cdot)$ is the indicator function. 
If the redundancy score exceeds a predefined threshold $\tau_{\text{redundancy}}$, the GPT model is re-prompted to generate new captions or modify existing ones to introduce more variety. The process is repeated until either full coverage is achieved ($V_{\text{final}}[b] = 1 \ \forall b \in B_{\text{custom}}$) or a maximum number of iterations, $I_{\text{max}}$, is reached to prevent the algorithm from running indefinitely in cases where certain aspects are difficult to cover.
The algorithm is described in Algorithm \ref{alg:E-COCAA}:
\begin{algorithm}[h]
\caption{Coverage-Driven Prompt Augmentation}
\label{alg:E-COCAA}
\textbf{Input:} $S_{\text{random}}$, $B_{\text{custom}}$, GPT-3.5-turbo, Maximum iterations $I_{\text{max}}$\\
\textbf{Output:} $S_{\text{final}}$ with full coverage of $B_{\text{custom}}$
\begin{enumerate}
    \item Initialize $S_{\text{final}} = S_{\text{random}}$. For each $t_i \in S_{\text{random}}$, compute $V_{t_i}$.
    \item Define the global coverage vector $V_{\text{final}}$ and identify all $b \in B_{\text{custom}}$ where $V_{\text{final}}[b] = 0$.
    \item For each uncovered aspect $b \in B_{\text{custom}}$, generate new caption $t_j$ using GPT: $t_j = G(\text{prompt}=b, \text{context}=S_{\text{final}})$. Add $t_j$ to $S_{\text{final}}$, and update $V_{\text{final}}$.
    \item Repeat until $V_{\text{final}}[b] = 1 \ \forall b \in B_{\text{custom}}$ or maximum iterations $I_{\text{max}}$ are reached.
    \item Compute the redundancy score $R(S_{\text{final}})$. If $R(S_{\text{final}}) > \tau_{\text{redundancy}}$, re-prompt GPT for more diversity.
\end{enumerate}
\end{algorithm}

\subsection{Activation-Aware Calibration Sampling Algorithm}
\label{sec:activation_calibration}
The activation-aware calibration sampling algorithm optimizes the selection of intermediates for calibration by prioritizing layers with higher activation variance, ensuring that the most informative layers are used for calibration. 
The novel calibration set generation method utilizes a GPT-based Caption Coverage algorithm to create a diverse set of prompts that cover various audio aspects and modalities. 
The Activation-Aware Calibration Sampling Algorithm selects timesteps where activation variance is highest, focusing on critical regions in the denoising process where quantization errors are more likely to propagate.
For each layer $l \in \{1, \dots, L\}$ of the diffusion model $\mathcal{M}$, we compute the activation variance $\sigma_l^2$. These variances are normalized to ensure that the probabilities sum to 1, with the normalized variance for each layer defined as $\hat{\sigma}_l^2 = \frac{\sigma_l^2}{\sum_{l=1}^L \sigma_l^2}$. 
The sampling probability for each layer is then computed as $P_l = \frac{\hat{\sigma}_l^2}{\sum_{j=1}^L \hat{\sigma}_j^2}$, ensuring proper normalization.
The threshold $\tau_{\text{var}}$ determines whether a layer is sufficiently informative for calibration. In practice, $\tau_{\text{var}}$ can be set empirically based on the distribution of variances across layers, often as a fraction (e.g., 10\%) of the maximum variance across all layers. Suppose no layer has variance exceeding $\tau_{\text{var}}$. In that case, the algorithm includes a fallback mechanism to gradually reduce $\tau_{\text{var}}$ until at least one layer qualifies, ensuring the process does not enter an infinite loop.
Let $z_i$ represent the intermediate sampled from layer $l$. The set of intermediates $S_{\text{calib}}$ forms the calibration set used to optimize the quantization process. This sampling procedure is repeated until the desired number of intermediates, $K$, is obtained.

\begin{algorithm}[h]
\caption{Activation-Aware Calibration Sampling}
\label{alg:activation-aware-calibration}
\textbf{Input:} Diffusion model $\mathcal{M}$, $L$ layers, Calibration set size $K$, Variance threshold $\tau_{\text{var}}$, Activation variances $\{\sigma_l^2\}_{l=1}^L$\\
\textbf{Output:} Set of intermediates $S_{\text{calib}}$
\begin{enumerate}
    \item Compute activation variances $\sigma_l^2$ for each layer $l$.
    \item Normalize variances: $\hat{\sigma}_l^2 = \frac{\sigma_l^2}{\sum_{l=1}^L \sigma_l^2}$.
    \item Define sampling probabilities $P_l$ for each layer: $P_l = \frac{\hat{\sigma}_l^2}{\sum_{j=1}^L \hat{\sigma}_j^2}$.
    \item Sample $K$ intermediates from layers according to $P_l$. Let $z_i$ be the intermediate sampled from layer $l$.
    \item For each sampled intermediate, check if $\sigma_l^2 \geq \tau_{\text{var}}$. If no layer satisfies the condition, iteratively reduce $\tau_{\text{var}}$ until at least one layer qualifies.
    \item Repeat until $|S_{\text{calib}}| = K$.
\end{enumerate}
\end{algorithm}


\begin{table*}[htp]
\centering
\caption{Main results for quantization across Tango, Make-An-Audio, and AudioLDM. Only non-sensitive layers are quantized and calibrated. Attention layers are at full precision. One hundred sampled prompts with 700 enhanced prompts were created for calibration.}
\label{tab:mainresults}
\resizebox{\textwidth}{!}{%
\begin{tabular}{|c|c|c|c|ccc|l|}
\hline
\multicolumn{1}{|l|}{\multirow{2}{*}{\textbf{Model}}} & \multicolumn{1}{l|}{\multirow{2}{*}{\textbf{\begin{tabular}[c]{@{}l@{}}U-Net \\ Parameters\end{tabular}}}} & \multicolumn{1}{l|}{\multirow{2}{*}{\textbf{Bitwidths(W/A)}}} & \multirow{2}{*}{\textbf{\begin{tabular}[c]{@{}l@{}}U-Net Size\\(Reduction\%)\end{tabular}}} & \multicolumn{3}{c|}{\textbf{Objective Metrics}} & {\multirow{2}{*}{\textbf{\begin{tabular}[c]{@{}l@{}}MOS-\\ ovl\end{tabular}}}} \\ \cline{5-7}
\multicolumn{1}{|l|}{} & \multicolumn{1}{l|}{} & \multicolumn{1}{l|}{} &  & \multicolumn{1}{l|}{\textbf{FD$\downarrow$}} & \multicolumn{1}{l|}{\textbf{FAD$\downarrow$}} & \textbf{KL$\downarrow$} &  \\ \hline
\multirow{3}{*}{Tango} & \multirow{5}{*}{866M} & Full Precision &-& \multicolumn{1}{l|}{24.52} & \multicolumn{1}{l|}{1.59} & 1.37 & 97.40 \\ \cline{3-8} 
 &  & 8W16A & 39\% & \multicolumn{1}{l|}{25.11} & \multicolumn{1}{l|}{1.59} & 1.38 &  98.16\\ \cline{3-8} 
 &  & 4W8A & 68\% & \multicolumn{1}{l|}{25.18} & \multicolumn{1}{l|}{1.68} & 1.38 &  97.20\\ \hline
\multirow{3}{*}{AudioLDM-L} & \multirow{5}{*}{739M} & Full Precision & -& \multicolumn{1}{l|}{27.12} & \multicolumn{1}{l|}{2.08} & 1.86 & 87.14 \\ \cline{3-8} 
 &  & 8W16A & 39\% & \multicolumn{1}{l|}{28.16} & \multicolumn{1}{l|}{2.17} & 1.91 &  90.36\\ \cline{3-8} 
 &  & 4W8A & 70\% & \multicolumn{1}{l|}{28.43} & \multicolumn{1}{l|}{2.43} & 2.17 &84.20  \\ \hline
 \multirow{3}{*}{Make An Audio} & \multirow{5}{*}{453M} & Full Precision &-& \multicolumn{1}{l|}{18.32} & \multicolumn{1}{l|}{2.66} & 1.61 &  93.24\\ \cline{3-8} 
 &  & 8W16A & 38\%& \multicolumn{1}{l|}{18.61} & \multicolumn{1}{l|}{2.78} & 1.64 & 97.48 \\ \cline{3-8} 
 &  & 4W8A & 62\%& \multicolumn{1}{l|}{18.94} & \multicolumn{1}{l|}{2.80} & 1.66 & 94.50 \\ \hline

\end{tabular}%
}
\end{table*}

\begin{figure}[ht]
    \centering
    \includegraphics[width=0.8\columnwidth]{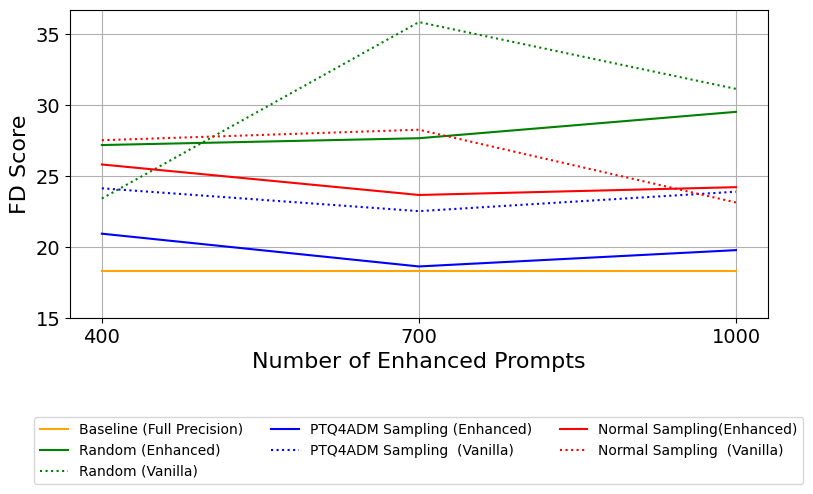}
    \caption{FD score of an 8W16A quantized Make-An-Audio model with various enhanced prompts based on 100 sampled captions from the AudioCaps dataset. The analysis includes intermediate sampling across timesteps- Random, Normal, and Ours along with enhanced prompts vs vanilla prompts from the dataset.}
    \label{fig:fd_score}
\end{figure}

\section{Evaluations}
\subsection{Evaluation Setup}
We evaluate Tango \cite{majumder2024tango}, Make-An-Audio \cite{huang2023make}, and AudioLDM \cite{liu2023audioldm} for text-conditional audio generation. Models are trained using original hyperparameters and datasets with frozen pre-trained text and audio encoders. We quantize weights and activations post-training.
\\ \textbf{Bitwidth configurations:} 32W/32A (baseline), 16W/16A, 8W/16A, 8W/8A, and 4W/8A, where 'W' denotes weights and 'A' denotes activations. These configurations explore trade-offs between compression, efficiency, and audio fidelity for resource-constrained environments.
\\ \textbf{Evaluation metrics} include Frechet Distance (FD), Frechet Audio Distance (FAD), and Kullback-Leibler (KL) divergence. Mean Opinion Score (MOS) is conducted with seven expert listeners, who rate 30 randomly sampled audio clips per configuration (0-100 scale) in a blind assessment and grade the overall perceptual quality of the generated audio.

\subsection{Experiments}
We focus on calibration set size, layer sensitivity, and bitwidth reduction, comparing PTQ4ADM-quantized models to full-precision (32W/32A) baselines. Results in Table \ref{tab:mainresults} show PTQ4ADM's exceptional performance, with PTQ4DM's sampling as a baseline (Figure \ref{fig:fd_score}). \textbf{Bitwidth reduction }to 8W/16A yields substantial model size reduction with minimal quality loss. A further decrease to 4W/8A introduces noticeable degradation but offers efficiency trade-offs (Table \ref{tab:mainresults}). We compare three \textbf{intermediates sampling strategies}: random (uniform across timesteps), normal (based on latent space mean, similar to PTQ4DM \cite{shang2023post}), and PTQ4ADM (prioritizing high activation variance timesteps). PTQ4ADM outperforms others, especially with enhanced prompts.
\textbf{Increasing enhanced prompts} based on AudioCaps dataset improves FD scores, with diminishing returns beyond 700 prompts (Figure \ref{fig:fd_score}). Layer sensitivity analysis confirms initial and final U-Net backbone layers are susceptible to low-bitwidth quantization, consistent with \cite{shang2023post}.

Our experiments demonstrate that strategic calibration and custom latent sampling significantly improve audio diffusion model robustness under low-bitwidth quantization. Lower bitwidths (e.g., 4W/8A) introduce more perceptual degradation, mitigated by our sampling strategy. Enhanced prompts improve FD scores compared to vanilla AudioCaps prompts (Figure \ref{fig:fd_score}).
PTQ4ADM's sampling strategy, focusing on high activation variance timesteps, outperforms random and normal sampling. While significantly improving efficiency, extreme quantization (4W/8A) can still degrade audio quality for complex scenes or detailed compositions.
PTQ4ADM maintains high audio fidelity while significantly reducing model size, making it suitable for resource-constrained applications. MOS maintains comparable quality across bitwidths and models, while human perception sometimes perceives the quantized model as better than the full precision model.

\section{Conclusions}

In this work, we presented PTQ4ADM, a novel post-training quantization method for text-conditional ADMs. 
Our approach, consisting of a GPT-based Caption Coverage Module and Activation-Aware Calibration Sampling Algorithm, achieved up to 70\% model size reduction with only minor increases in FAD, FD and KL scores. 
This work enhances the deployment of high-quality audio models in resource-limited settings, enabling on-device audio creation. Future efforts could focus on adaptive quantization and perceptual loss functions to better balance quality and efficiency.


\end{document}